# Artificial Tikkun Olam:

## AI Can Be Our Best Friend in Building an Open Human-Computer Society

Simon Kasif

Department of Biomedical Engineering

Boston University,

Boston, MA, USA

* Correspondence should be sent to kasif@bu.edu

**Abstract**

Technological advances of virtually every kind pose risks to society including fairness and bias. We review a long-standing wisdom that a widespread practical deployment of any technology may produce "adverse side effects" misusing the knowhow. This includes AI but AI systems are not solely responsible for societal risks. We describe some of the common and AI specific risks in health industries and other sectors and propose both broad and specific solutions. Each technology requires very specialized and informed tracking, monitoring and creative solutions. We postulate that AI systems are uniquely poised to produce conceptual and methodological solutions to both fairness and bias in automated decision-making systems. We propose a simple "intelligent system quotient" that may correspond to their adverse societal impact and outline a multi-tier architecture for producing solutions of increasing complexity to these risks. We also propose that universities may consider forming interdisciplinary Study of Future Technology Centers to investigate and predict the fuller range of risks posed by technology and seek both common and AI specific solutions using computational, technical, conceptual and ethical thinking and analysis.

**Introduction**

Artificial Intelligence (AI) [1-6] has been recently dominating the technology news by demonstrating remarkable practical advances and applications across many domains. But in parallel, it also created an avalanche of critiques both inside and outside the field creating anxiety and perhaps even fear of AI and the risks presents. The analysis of this topic from experts in AI usually address specific technical, theoretical or ethical issues and risks that are indeed receiving increased attention from thousands of trained AI scientists and engineers. However, the accounts received from external sources are routinely narrowly informed and often conflate AI specific risks and concerns with broad challenges that are shared by many technologies and Data Science in general and include many predictive methods such that Statistics, Epidemiology, Economics, Behavioral Sciences, Biomedical Informatics, Systems Engineering, Organizational Science and more. In this conceptual perspective, we aim to present a broad analysis of this topic hoping to produce a clearer, birds eye view of AI and future risks it may evolve to produce. We begin with an a repeated but not particularly broadly informed critique of AI, singling out the bias exhibited by machine learning systems exemplified by the "POV: Artificial Intelligence Has a Powerful Brain, but It Still Needs a Heart":

"That problem is twofold: there are the underpinnings of technology itself, and there is the application of AI in ethical and unbiased ways. Machines may be fast learners, but the data they learn from are often a compilation of human decisions. If those decisions are freighted with bias, and they often are, AI can bake that in too, creating a system that is perpetually unfair. Research has shown, for example, that some AI-powered facial recognition software returns more false matches for African Americans than it does for white people. Such technological shortcomings are exacerbated by AI's failure to recognize them as shortcomings, as humans might do. Algorithms do not self-correct. They self-reinforce", Azer Bestavros, Boston University.

We generally agree with this sentiment and the need for a broad discussion on human rights in a fair and open society [7]. Discrimination should not be tolerated in neither society nor computer systems [8] [9, 10]. However, we feel that much is missed in these narrowly targeted perspectives that single out AI as the sole perpetrator of bias. The popular press picks up on these views and the fear of AI is amplified in the popular media. Fear is rarely a good impetus for long term rational solutions.  Here we will argue that the current fairness, bias and inadequate risk assessment concerns are important but are not exclusive to AI and are broadly applicable to most of data science and related predictive methodologies. Thus, we **MUST** separate broadly applicable concerns that require general and common solutions (such as ethics, causality, experiment design, trust in computers or quality of data) and AI specific challenges such as reasoning about knowledge, artificial culture, collective intelligence or evolution of a super-intelligence. A helmet might be a good solution for a person riding on a bike or a passenger on a rocket to space. But there are many unique challenges, risks and complexity associated with advanced AI technologies. The heterogeneity of intelligent systems, technologies and risks require a broad philosophical, scientific and engineering approach to repairing the world (Tikkun Olam) defined by a society of humans and intelligent systems.

**AI Systems are Complex, Rapidly Evolving and Difficult to Anticipate**

In order to fully anticipate and prevent AI risks we need to have some appreciation for the broad goals of AI and a vision that can anticipate emerging challenges.  The rapid progress in AI requires a big picture of the field in order to predict risks and produce solutions.  In Figure 1 we tried to visualize the AI field as a multidimensional hypercube that captures several of the many dimensions that define AI.  We placed many of the major AI landmarks on just a bounded region of a single plane demonstrating the big picture, imagination and technical prowess required to make progress in AI and anticipate future risks. bigger picture. The range of applications of intelligent systems is virtually unbounded, thus the problem of AI safety and bias requires a broader examination of the common and specific risks across the many modalities of the field.

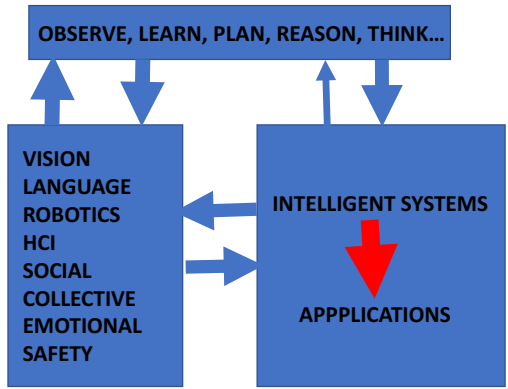 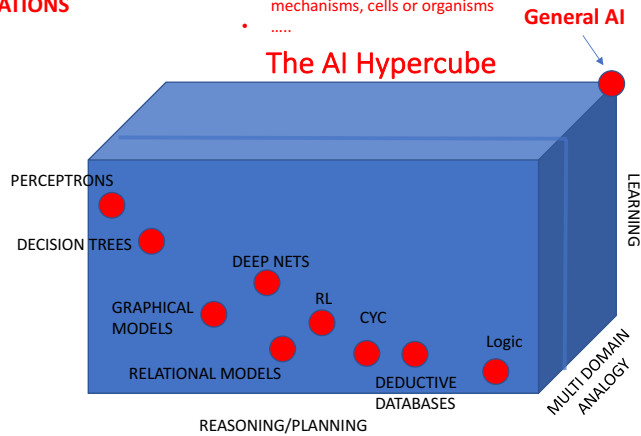

Figure 1: The AI Hypercube: From Simple Neural Nets and Deductive Databases to General AI

References: Perceptrons, Decision Trees, Deep Nets [1-6] [11], Graphical Models and Relational Models [12-15], Deductive Databases, Logic, Cyc [16]

**AI is not alone in creating bias and fairness challenges**

We first note that rarely do we see scholars accuse a t-test or other textbook statistical procedures of lack of soul or absence of heart. Because AI in its ultimate vision is aimed to mimic human intelligence, it is tempting to associate AI algorithms with "human traits" like a "mind', "heart" or "empathy" [17]. However, non-specialists struggle to isolate AI specific challenges and more basic problems in data and predictive sciences that are shared with other fields. In many cases it is not the AI technology that is to blame. The real challenges hide in the experimental design, lack of proper treatment of causality, confounders, human errors and many others logical, statistical and logistic problems and trade-offs. These challenges are intricately shared between AI and other predictive inference algorithms deployed universally. Moreover, these issues are fundamental in "establishing trust in networks of computers and humans" [15, 18, 19] and human centered computing [10]. The complexity of these interdisciplinary challenges and their significant cross-overs in evolving AI systems is summarized in Figure 2.

Virtually every time we use a statistical inference or a predictive classification, the outcome might be biased by the data used in the study or implicit design decisions. Even if it's a simple test used routinely to compare the means of populations or comparing individuals to these means. For instance, an admission to college is based in part on comparing the GPA or a GRE score of an applicant with the mean of admitted students. If these scores are lower in some population an implicit bias is created.  Every time we apply for a loan or insurance coverage, textbook statistical procedures are used in evaluating our applications.  Similar issues of fairness, relative risk and bias arise. As one example, a geographic location can be easily disguised as a parameter that might discriminate against a particular race or socio-economic slice. Insurance companies routinely assess risk and make policy decisions based on data that may include the age, gender or employment status of applicants producing implicit biases. The Affordable Care Act partially addressed discrimination issues found in U.S. health insurance systems and introduced improved anti-discrimination coverage and provisions.  However, many challenges remain. In a different discourse, a statistical algorithm called COMPAS was used to evaluate the risk a criminal defendant might commit a crime after release. Here, a hidden from public view statistic was deployed to keep human beings jailed. As a final example from the world of cancer medicine, the NPI statistical score has been used in routine clinical to predict prognosis of a breast cancer patient and guide treatment.  None of these methods use AI, however, all inherently biased by the distribution of data used to derive these statistics.

Now, let's consider another area where data science is routinely used for critical life and death decisions.  A clinical test (diagnostic) or a drug are typically tested on a specific population. There is no AI, just textbook statistics so far but these clinical decisions are inherently biased by the patient cohorts selected for the trial, confounders or proper treatment of causality.  While the design of these cohorts aims to be randomized, it's difficult to completely purge data bias. For instance, clinical trials often choose to eliminate patients with independent health issues but in practice such patients are obviously more common than not in real clinical settings. It's actually becoming even further complicated with precision medicine, as cohorts are increasingly biased towards successful trials on genetically targeted populations. The book "Overdosed America" by Johns Abramson documents the fact that Pharmaceutical companies often conduct clinical trials on adult men but the drugs are prescribed much more broadly and might include off label use in women. One growing concern is the off-label use of anti-depressants in the pediatric population without robust evidence of effectiveness in this population as they are typically clinically tested on adults only [20, 21].  Thus, these clinical decisions are guided by simple statistical inferences derived from biased cohorts in a similar fashion to the bias complex machine learning programs may produce if they are trained on small and biased (by design) data that does not represent the general clinically targeted population.

The majority of genomic DNA reports we receive today are not based on an all-encompassing catalogue of race specific genomes and are likely to produce increased error in under-represented populations [22]. These reports are produced by algorithms using textbook statistics. No AI so far.  The situation is significantly more complex in epidemiological human studies over large populations where it is even more challenging to obtain accurate recommendations due to bias, confounders, missing data and the challenges of dealing with causality. We recently saw a bit of that in the public debate about whether red meat causes cardiovascular events following one large but controversial study that appeared to claim otherwise.

This brings us to the broad role of causality [23] in addressing ethnical computing generally and AI specifically. Gender discrimination should be eliminated in any form. Indeed, AI programs that "learn" from text may start exhibiting biases already engrained in the data such as a homemaker must be a woman and a pilot must be a man. This bias should be tracked, documented and corrected. However, AI does not explain the salary disparity in the pay of American soccer players which is human made, conceivably due to a flawed economic or behavioral marketing analysis. A recent study on gender discrimination in NIH reviews showed that women are receiving lower scores than men. However, a deeper examination illustrates an important confounder, namely the emphasis in the grant. It is theoretically possible that the applications submitted by women scientists were enriched in patient centered research vs male applications that were more focused on more technical and less translational work. It is challenging to de-convolve these variables and it is conceivable that both patient centered and applications from females are scored more harshly in panel reviews. Interventions could help to untangle this complexity [23].  Similar to data bias, explicit or implicit preferences, and other fundamental issues, **causality is at the core** of evaluating fairness and is broadly applicable.

To reiterate many of the concerns about bias, fairness, risk and utilities in computerized decision and predictive modeling systems are not exclusive to AI.  The conceptual issues touch on fundamental topics in decision theory, rationality, expected utilities, subjective probabilities, causality and theory of preferences and their applications in forecasting or prediction [10, 24-26], theory and evolution of cooperation [27-29] and philosophy of morality.  The practical issues are equally challenging and involve basic ethics and right and wrong reasoning. E.g. balancing the important need to deliver a life-saving drug to a needy population and conducting a prohibitively costly large clinical trial.  The **"incidentalome",** or the trade-off to detect a disease early and raising false alarm resulting in costly follow-ups is a common challenge in defensive medicine.  A most dramatic illustration of the trade-offs and biases in predictive sciences is the apparent underestimate of how fast global climate changes are taking place, the causal role of human made impact and the enormous risks posed by not responding to it in a timely fashion.

**History and Unique AI Risks**

A historical perspective may be insightful as well.  Prof. Joseph Weizenbaum at MIT was an early crusader, warning us of the risks posed by AI more than 40 years ago. He was the famous father of Elisa (a rudimentary AI algorithm that mimics a therapist). But Prof. Weizenbaum was thinking broadly about **all** technologies that produced unintended consequences and risks and he feared AI will follow the same course. Today this broad list may include air pollution, misuse of opioids, food industry, the internet and many more.

Norbert Wiener was another pioneer in AI sounding off early concerns about automation. Systems engineers are aware of the fact that the simple thermostat we use in our homes to control temperature or glucose sensors used for diabetes management are evolutionary ancestors of much more complex modern control and AI systems deployed in robots for surgical planning or clinical trials. The conceptual multi-layer architecture in modern deep learning systems [30] can be ascertained from early systems theory [31]. Of course, it took great technical genius, inspiration and heroic perspiration to scale it up to modern deep learning capabilities. But any automation and AI share risks independent of whether they respond to a simple sensor, computer vision or natural language.  In recognition of the early developments in automation and system engineering which were subsequently scaled up and generalized by AI scientists to complex, general and highly scalable AI methods (see supplement) we refer to our informal goal as a **"heart filter"**.

There are indeed, technical and logical challenges that are specific to AI software.  At the technical level, lack of interpretability in deep AI systems stirred many developments to make AI more explainable. These efforts towards greater transparency are found in many AI laboratories but we will just mention the recent emphasis on transparency by Prof. Yoshua Bengio and his group, the long standing emphasis on causality by Prof. Judea Pearl at UCLA and the programmable AI frameworks at Berkeley among others [12-15].

However, if one actually thinks imaginatively, AI also poses long term risks that are unique to the field. For instance, consider the eventual risk AI poses for education. Eventually, AI will develop a transformative and useful capability to write assays and perhaps even books. But imagine students in schools using this natural language technology in the same way they use Mathematica to do college level math today. There would be no simple way to detect it.

AI has an exponentially increasing role in biological discovery [32-36] producing a long term prohibitive challenge in tracing predictions to their experimental evidence, separating computational hypotheses and facts. Similarly, consider the more radical proposals and emerging trend to use AI to guide scientific inquiry as described in the COMBREX (Computational Bridges to Experiments) PROJECT and Robot Scientist [37, 38]. These proposals go back to the founders of AI and will create their own challenges in prioritizing semi-automated scientific research.

While much attention has been directed to the biases that may affect specific minorities, less work has been done on small groups such as people with disabilities. Speech recognition or machine translation systems used by the speaking impaired and Parkinson's or Alzheimer's patients might make critical errors resulting in accidents.

There is an inherent **cultural risk** associated with deep learning and other black box methods that take in big data and produce solutions. These techniques do not require deep domain expertise and over time can make all of us intellectually lazy and have a profound impact on science, medicine, engineering and culture. My old colleague at Johns Hopkins Fred Jelinek was one of the fathers of statistical speech and language processing. His group at IBM included a number of luminaries that pioneered the early ideas that revolutionized practical speech systems leading to Alexa and Siri and subsequently the almost inconceivable rapid progress in practical machine translation. Fred used to say "Every time I fire a linguist, the performance of my speech system goes up". A deeply profound and disturbing prediction of the impact AI black boxes could have on culture and science.

There are hundreds of similarly imaginative applications of AI that may create significant and unique challenges that are not relevant to other fields. These specialized anomalies created by sophisticated AI systems should not be conflated with more rudimentary bias and fairness issues and challenges that must be studied and combatted broadly (Figure 2).

**Solutions: AI can be our best friend in implementing the "Heart Filter"**

First and foremost, we note that AI software and methodologies can actually be deployed effectively to reduce biases created by cultural or technological practices. This idea can be implemented by training AI algorithms to recognize bias or using high level AI languages based on logic or other fair representations to program transparent and fair decisions or allocation of resources. We actually expect AI coupled with community science [37, 39] to play a significant role in reducing or mitigating some of the risks by appropriately monitoring human and technological activities, improving compliance and early identification of anomalies, producing timely reports and planning algorithms with quantified expected utilities or built in preference logics. We can already see the early seeds of such AI software developed to detect profiling, discover problematic behaviors [40], improve clinical processes in cancer and more [18, 41-48]. At the more fundamental level, we expect causal networks and causality based inference methods to play an increasing role in implementing fairness in predictive systems including AI.

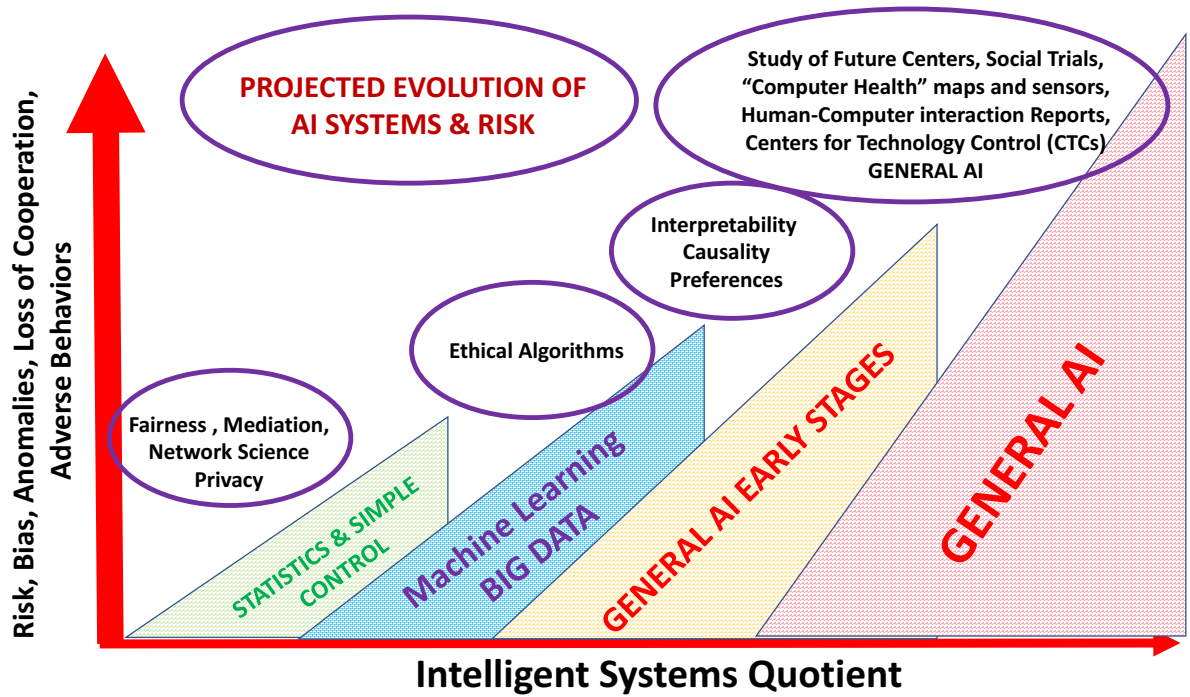

**Figure 2:** The Evolution of AI Systems, Adverse Impact and Possible Solutions for Tikkun Olam

      In summary, the "Heart Filter" should be applied to every predictive or decision science and related technologies not just AI. Technological advances of virtually every kind pose risks to society. Both cars and planes can be turned into destructive weapons without proper control and oversight. Seat belts, speed limits and automation (including AI based navigation) are solutions for risks created by both technologies. But each technology requires very specialized and informed tracking, monitoring and creative solutions. Universities may consider forming interdisciplinary Study of Future Technology Centers to investigate and predict the fuller range of risks and seek both common and AI specific solutions using computational and ethical thinking. These centers should identify key conceptual, ethical, technical and engineering challenges. The study of causality has demonstrated a multi-disciplinary impact at all four levels. Network science is another example of common issues and includes topics such as studying influence in networks [49-52] and network propagation methods [53, 54] . Ultimately, a widespread practical deployment of any technology might produce "adverse side effects" misusing the knowhow. The recent developments related to false advertisement on social networks is one such example. We may need to start designing appropriate "social trials" in analogy to clinical trials to test technologies on selected cohorts that are best in assessing comparative effectiveness and adverse effects. We may need Consumer Reports style solutions or we may even consider forming an agency similar to CDC (e.g., Center for Technology Control (CTC)) that will provide both detection and rapid response to any rapidly growing risks produced by predictive technologies. In analogy to mobile health, intelligent sensor networks [55, 56] and "computer-health" maps [57, 58] may provide a platform for early detection of anomalous human-computer behaviors. However, we remain

optimistic about the role of AI in an open and ethical society of humans and computers [15, 42, 51]. Figure 2 summarizes some of these proposed solutions

We close by an old but still relevant insight stipulated the philosopher of science Carl Popper: **"To conclude, Popper's theory of knowledge and of science seems to support the view that in order to do good science, we need also to become a specific kind of person, and hence, being a scientist is not just a logical or epistemological endeavor, but ethical as well."** Emanuele Ratti (see supplement). This recommendation should apply to both humans and computers.

**Acknowledgement:**  We are deeply grateful to colleagues and friends who provided feedback on early versions or influenced (in old discussions over the years) the ideas expressed in this perspective. In particular, Drs. Robert Berwick, Art Delcher, Charles Cantor, Jim Collins, Charles DeLisi, James Flanagan, Tim Gardner, Russ Greiner, Tom Huang, Ron Kahn, Adi Karni, Isaac Kohane, Stan Letovsky, Jack Minker, Judea Pearl, Rich Roberts, Azriel Rosenfeld, Stuart Russell, Peter Salovey, Peter Szolovitz and David Waltz.  The specific views expressed here are the sole responsibility of the author.

# SUPPLEMENT

We provide selected recommendations for both broad and highly technical readings sorted by topics. The recommendations vary in technical difficulty from popular articles to complex mathematical analysis.  The readings are all over the map by design to illustrate the common challenges across disciplines and the solutions offered by both AI and General Data Science.  We obviously cannot cover all the relevant knowledge.  There are many subareas in AI that emphasize development or learning of logical, probabilistic and other ways to code declarative or procedural knowledge such as medical ontologies.   There are innumerable AI applications in areas such as machine translation, finance, commerce, industrial robotics, information retrieval and summarization, web-search and more.  One area we plan to explore in more detail in the future is the important issue of AI and tracing provenance and causal explanations.  We hope this perspective allows to sample a few landmarks and elucidate both common and specialized challenges in confronting adverse consequences in building and deploying AI systems and contrasting them with common problems that emerge in statistical and decisions sciences.

**General, Popular Press and Policy Discussion on Ethics, AI and Technology**

https://president.yale.edu/speeches-writings/speeches/repair-world

https://www.sciencedaily.com/releases/2019/11/191107160658.htm

https://ctshf.nd.edu/posts/popper-and-the-ethical-dimension-of-the-scientific-method/

https://www.washingtonpost.com/opinions/its-not-up-to-mark-zuckerberg-to-decide-what-news-is-legitimate/2019/11/07/2f447ed0-01a5-11ea-8bab-0fc209e065a8_story.html

https://www.nytimes.com/2018/02/12/business/computer-science-ethics-courses.html

https://www.businessinsider.com/microsoft-gab-azure-cloud-anti-semitism-2018-8

https://www.theguardian.com/world/2016/mar/29/microsoft-tay-tweets-antisemitic-racism

https://en.wikipedia.org/wiki/Effective_altruism

https://futureoflife.org/2019/08/14/how-can-ai-systems-understand-human-values/

**AI: Popular but Thought Provoking**

Ray Kurzweil, The Singularity Is Near: When Humans Transcend Biology, Penguin Books, 2006.

Gerald Edelman and Giulio Tononi, A Universe of Consciousness: How Matter Becomes Imagination, Basic Books, 2000, Reprint edition 2001. ISBN 0-465-01377-5

Churchland, P. S. and Sejnowski, T. J., The Computational Brain, MIT Press 1992.

Terrence J. Sejnowski, The Deep Learning Revolution, MIT Press. 2018

CYC by Doug Lenat   https://en.wikipedia.org/wiki/Cyc

Judea Pearl and Dana Mackenzie, The Book of Why: The New Science of Cause and Effect New York: Basic Books, 2018

## Ethics

Ethics in the Real World: 82 Brief Essays on Things That Matter, Peter Singer, Princeton University Press, 2016

Emanuele Ratti,   Popper and the Ethical Dimension of the Scientific Method. Blog, 2017
https://ctshf.nd.edu/posts/popper-and-the-ethical-dimension-of-the-scientific-method/

Karl R. Popper (1950), The Open Society and Its Enemies, Vol. I & II (Princeton, NJ: Princeton University Press), pp. 570–571.

Johnson, Robert (6 April 2008) [23 February 2004]. "Kant's Moral Philosophy". Stanford Encyclopedia of Philosophy. April 2012.

## Scientific methods, Fairness, Cooperation, Error, Risk and Preferences

Fairness and the assumptions of economics
D Kahneman, JL Knetsch, RH Thaler The Journal of Business Vol. 59, No. 4, Part 2: The Behavioral Foundations of Economic Theory (Oct., 1986), pp. S285-S300

Preference Logics:
https://plato.stanford.edu/entries/preferences/#PreLog

Simon, Herbert A, 1986. "Rationality in Psychology and Economics," The Journal of Business, University of Chicago Press, vol. 59(4), pages 209-224, October.

Axelrod, Robert, The Evolution of Cooperation (Revised ed.), Perseus Books Group, (2006),

Ioannidis, J. P. A. (2005). "Why Most Published Research Findings Are False". PLoS Medicine. 2 (8): e124. doi:10.1371/journal.pmed.0020124. PMC 1182327. PMID 16060722.

Sigmund, Karl; Fehr, Ernest; Nowak, Martin A. "The Economics of Fair Play", Scientific American, vol. 286 no. 1, pp. 82–87, (January 2002).

Kleinberg J, Lakkaraju H, Leskovec J, Ludwig J, Mullainathan S., Human Decisions and Machine, Predictions, Q J Econ. 2018 Feb 1;133(1):237-293. doi: 10.1093/qje/qjx032. Epub 2017 Aug 26., PMID: 29755141

D Lupton, Risk and sociocultural theory: New directions and perspectives
Cambridge University Press, 1999

Rediet Abebe, Solon Barocas, Jon Kleinberg, Karen Levy, Manish Raghavan, David G Robinson Roles for computing in social change, Proceedings of the 2020 Conference on Fairness, Accountability, and Transparency Pages 252-260, 2020.

## Discrimination in Insurance Industry

How We Examined Racial Discrimination in Auto Insurance Prices Jeff Larson, Julia Angwin, Lauren Kirchner, Surya Mattu for Pro Publica and
Dina Haner, Michael Saccucci, Keith Newsom-Stewart, Andrew Cohen, Martin Romm for Consumer Reports, April 5, 2017
https://www.propublica.org/article/minority-neighborhoods-higher-car-insurance-premiums-methodology

## AI, Agents, Networks, Behavior and Ethical Reasoning

Norbert Wiener, Some Moral and Technical Consequences of Automation, Science, 06 May 1960: Vol. 131, Issue 3410, pp. 1355-1358

AI Open Letter
https://futureoflife.org/ai-open-letter/

Philip S. Thomas1, Bruno Castro da Silva, Andrew G. Barto, Stephen Giguere, Yuriy Brun, Emma Brunskill, Preventing undesirable behavior of intelligent machines, Science 22 Nov 2019, Vol. 366, Issue 6468, pp. 999-1004

https://news.stanford.edu/2019/11/21/stanford-helps-train-ai-not-misbehave/

Stuart Russell, Man Compatible: Artificial Intelligence and the Problem of Control 2019.

Human Compatible AI
https://humancompatible.ai/

Y Hswen, K Sewalk, G Tuli, J Brownstein, J Hawkins, Using big data analytics and machine learning to assess racial disparities in patient experience, APHA's 2019 Annual Meeting and Expo (Nov. 2-Nov. 6)

Magua W, Zhu X, Bhattacharya A, Filut A, Potvien A, Leatherberry R, Lee YG, Jens M, Malikireddy D, Carnes M, Kaatz A., Are Female Applicants Disadvantaged in National Institutes of Health Peer Review? Combining Algorithmic Text Mining and Qualitative Methods to Detect Evaluative Differences in R01 Reviewers' Critiques. J Womens Health (Larchmt). 2017 May;26(5):560-570. doi: 10.1089/jwh.2016.6021. Epub 2017 Mar 10. PMID: 28281870

Rossi, F., Venable, K., Walsh, T.: A short introduction to preferences., Synthesis Lectures on Artificial Intelligence and Machine Learning. Morgan & Claypool (2011)

A case-based approach to modeling legal expertise
KD Ashley, EL Rissland - IEEE Intelligent Systems, 1988

https://intelligence.org/files/EthicsofAI.pdf

https://www.cnbc.com/2018/03/14/allen-institute-ceo-says-a-i-graduates-should-take-oath.html

Yuan Y, Alabdulkareem A, Pentland A'., An interpretable approach for social network formation among heterogeneous agents. Nat Commun. 2018 Nov 8;9(1):4704. doi: 10.1038/s41467-018-07089-x. PMID: 30410019

Woolley AW, Chabris CF, Pentland A, Hashmi N, Malone TW., Evidence for a collective intelligence factor in the performance of human groups. Science. 2010 Oct 29;330(6004):686-8. doi: 10.1126/science.1193147. Epub 2010 Sep 30. PMID: 20929725

**Causality**

Pearl, J., Glymour, M., and Jewell, N. Causal Inference in Statistics: Primer. Wiley, 2016.

Pearl, Judea. Causal inference in statistics: An overview. Statistics Surveys, 3:96–146, 2009. 2

Galea S, Hernán MA., Win-win: Reconciling Social Epidemiology and Causal Inference. Am J Epidemiol. 2019 Oct 3. pii: kwz158. doi: 10.1093/aje/kwz158. PMID: 31579911

Glymour C, Zhang K, Spirtes P., Review of Causal Discovery Methods Based on Graphical Models. Front Genet. 2019 Jun 4;10:524. doi: 10.3389/fgene. 2019.00524. eCollection 2019. Review. PMID: 31214249

**AI and Medicine**

**Simple Controllers**:

Farmer TG Jr, Edgar TF, Peppas NA. The future of open- and closed-loop insulin delivery systems., J Pharm Pharmacol. 2008 Jan;60(1):1-13. Review. PMID: 18088499

Russell SJ, El-Khatib FH, Sinha M, Magyar KL, McKeon K, Goergen LG, Balliro C, Hillard MA, Nathan DM, Damiano ER., Outpatient glycemic control with a bionic pancreas in type 1 diabetes, N Engl J Med. 2014 Jul 24;371(4):313-325.

**AI:**

Yu VL, Buchanan BG, Shortliffe EH, Wraith SM, Davis R, Scott AC, Cohen SN., Evaluating the performance of a computer-based consultant. Comput Programs Biomed. 1979 Jan;9(1):95-102. PMID: 365439

Hinton G. Deep Learning-A Technology With the Potential to Transform Health Care. JAMA. 2018 Sep 18;320(11):1101-1102. doi: 10.1001/jama.2018.11100. No abstract available. PMID: 30178065

## Theory

**Network Influence:**

Kleinberg J., Analysis of large-scale social and information networks. Philos Trans A Math Phys Eng Sci. 2013 Feb 18;371(1987):20120378. doi: 10.1098/rsta.2012.0378. Print 2013 Mar 28. PMID: 23419847

David Kempe, Jon Kleinberg, and Éva Tardos. 2003. Maximizing the spread of influence through a social network. In Proceedings of the ninth ACM SIGKDD international conference on Knowledge discovery and data mining (KDD '03). ACM, New York, NY, USA, 137-146

**Common knowledge:**

Aumann, Robert (1976) "Agreeing to Disagree" Annals of Statistics 4(6): 1236–1239.

Fagin, Ronald; Halpern, Joseph; Moses, Yoram; Vardi, Moshe (2003). Reasoning about Knowledge. Cambridge: MIT Press

**Prediction Challenges in Statistics and ML**

Avrim Blum and Tom Mitchell. 1998. Combining labeled and unlabeled data with co-training. In Proceedings of the eleventh annual conference on Computational learning theory (COLT' 98). ACM, New York, NY, USA, 92-100.

Ioannidis J.P., Why most published research findings are false: PLoS Med. 2007

**Algorithmic Fairness**

Dwork, Cynthia, Hardt, Moritz, Pitassi, Toniann, Reingold, Omer, and Zemel, Richard. Fairness through awareness. In Proceedings of the 3rd Innovations in Theoretical Computer Science Conference, pp. 214–226. ACM, 2012.

Kilbertus, N., Carulla, M. R., Parascandolo, G., Hardt, M., Janzing, D., and Schölkopf, B. Avoiding discrimination through causal reasoning. Advances in Neural Information Processing Systems 30, 2017.

Kusner, M. J., Loftus, J., Russell, C., and Silva, R. (2017). Counterfactual fairness. In Advances in Neural Information Processing Systems, pages 4066–4076.

Sam Corbett-Davies and Sharad Goel, The Measure and Mismeasure of Fairness: A Critical Review of Fair Machine Learning, arXiv:1808.00023, 2018.

Simon S. Haykin, Kalman Filtering and Neural Networks, John Wiley & Sons, Inc. New York, NY, USA, 2001.

Abraham Othman, Christos H. Papadimitriou, Aviad Rubinstein:
The Complexity of Fairness Through Equilibrium. ACM Trans. Economics and Comput. 4(4): 20:1-20:19 (2016)

## Social Science, Big Data, Digital Health Maps and Mobile Sensors

Deborah Lupton, Self-tracking cultures: towards a sociology of personal informatics 2014/12/2 Proceedings of the 26th Australian Computer-human interaction conference on designing futures: The future of design, Pages, 77-86, ACM

Michael M, Lupton D., Toward a manifesto for the 'public understanding of big data', Public Underst Sci. 2016 Jan;25(1):104-16. doi: 10.1177/0963662515609005. Epub 2015 Oct 13. PMID: 26468128

D Lupton, Digital sociology, Routledge, 2014
Digital disease detection--harnessing the Web for public health surveillance.

Brownstein JS, Freifeld CC, Madoff LC., N Engl J Med. 2009 May 21;360(21):2153-5, 2157. doi: 10.1056/NEJMp0900702. Epub 2009 May 7. PMID: 19423867

Estrin D, Sim I., Health care delivery. Open mHealth architecture: an engine for health care innovation., Science. 2010 Nov 5;330(6005):759-60. doi: 10.1126/science.1196187. PMID: 21051617